\pdfoutput=1
\documentclass[aps,pre,superscriptaddress,amsmath,amssymb,reprint,nofootinbib,twoside]{revtex4-2}
\usepackage{graphicx}

\begin{document}

\title{Using correlation diagrams to study the vibrational spectrum of highly nonlinear floppy molecules: The K-CN case}
\author{H. P\'arraga}
\author{F. J. Arranz}
\email{fj.arranz@upm.es}
\author{R. M. Benito}
\email{rosamaria.benito@upm.es}
\affiliation{Grupo de Sistemas Complejos, Universidad Polit\'ecnica de Madrid,
28040 Madrid, Spain.}
\author{F. Borondo}
\email{f.borondo@uam.es}
\affiliation{Instituto de Ciencias Matem\'aticas (ICMAT), 
Cantoblanco--28049 Madrid, Spain.}
\affiliation{Departamento de Qu\'\i mica, Universidad Aut\'onoma de Madrid, 
Cantoblanco--28049 Madrid, Spain.}
\date{\today}

\begin{abstract}
The correlation diagrams of vibrational energy levels considering the Planck constant as a variable parameter have proven as a very useful tool to study vibrational molecular states, and more specifically in relation to the quantum manifestations of chaos in such dynamical systems. In this paper, we consider the highly nonlinear K-CN molecule, showing how the regular classical structures, i.e., Kolmogorov-Arnold-Moser tori, existing in the mixed classical phase space appear in the quantum levels correlation diagram as emerging diabatic states, something that remains hidden when only the actual value of the Planck constant is considered. Additionally, a quantum transition from order to chaos is unveiled with the aid of these correlation diagrams, where it appears as a frontier of scarred functions. 
\end{abstract}

\maketitle

\section{\label{sec:intro}Introduction}

The use of correlation diagrams, very often done in connection with the symmetry of the molecule~\cite{IachelloLevine}, to rationalize the knowledge on molecular rovibrational states has a long tradition in Molecular Spectroscopy; see for example~\cite{KellmanRice,Bunker}. Correlations diagrams have also been used in the study of molecular orbitals~\cite{McQuarrie}, inelastic collisions~\cite{Rice}, electronic states~\cite{Sharp}, or chemical reactivity~\cite{WoodwardHoffmann}, as well.

\vspace{2ex}   

Most often, correlation diagrams use a real magnitude as the varying parameter, such as geometrical distances or angles. However, nothing prevents the use of other more daring alternatives. In particular, our group has been using the value of the Planck constant, $\hbar$, artificially taken as a variable parameter, to elucidate the dynamical characteristics of vibrational states of floppy triatomic molecules~\cite{Arranz.LiCN-HCN-HO2.correlation.diagram}, and more specifically in relation to the quantum manifestations of chaos~\cite{Haake} in such dynamical systems.

\vspace{2ex}   

The idea is fairly simple, and it is based on the semiclassical argument put forward by Weyl~\cite{Weyl}, according to which a quantum state spans a volume in the system $N$-dimensional phase space proportional to $\hbar^N$, being the constant of proportionality a linear function of the quantum numbers. In this way, by making Planck's constant $\hbar \rightarrow 0$ we can force the quantum states to ``accommodate'' in smaller and smaller phase space volumes, being them confined after some point of this process into the regular classical region of the molecular phase space, where the dynamical characteristics of the quantum state can be easily ascertained. In other words, using $\hbar$ as a parameter represents an ideal tool to implements a kind of microscope that focuses with varying resolution on the classical regular structures embedded in chaotic regions.

\vspace{2ex}   

The method is fully explained in Sec.~\ref{sec:adiabatic.correlation.diagram}. In connection to this, some aspects have to be taken into account.

\vspace{2ex}   

Classically, the structure of the phase space associated to molecular vibrations is well understood in terms of the ideas of nonlinear dynamics~\cite{LL}. From this perspective, molecules can be viewed as Hamiltonian systems formed by collections of coupled nonlinear oscillators whose dynamics is well explained by the celebrated Kolmogorov-Arnold-Moser (KAM) and Poincar\'e-Birkhoff theorems~\cite{Berry}. At low energies, the intramolecular motion takes place around the stable equilibrium geometrical configurations of the molecule. The harmonic approximation is then valid, the Hamiltonian is very close to separability, the dynamics is regular (normal modes), and the trajectories are confined into $N$-dimensional invariant tori. As the excitation increases, the KAM dictates that some invariant tori (those with ``less irrational'' frequency ratios~\cite{Berry}) are destroyed, this giving rise to bands of stochasticity, bounded by the distorted surviving tori. These bands are ergodically explored by the corresponding trajectories. At these values of the energy the coupling among normal modes begins to be important, this marking a progressive transition to local modes dynamics~\cite{Sibert}. Prominent in the set of destroyed tori are the resonant ones, in which zeroth order commensurate relations among frequencies exist. The KAM theorem states that they are destroyed by minimal perturbations, and their fate is dictated by the Poincar\'e-Birkhoff theorem. According to it, an even number of periodic orbits (PO) survive the destruction~\cite{Berry}. Half of them are stable, and the other half unstable. Around the former stable motion takes place, which is organized in resonant invariant tori. In phase space these structures give rise to chain of islands, where much intramolecular vibrational energy transfer takes place~\cite{Sibert}. The latter originate chaos which is nevertheless organized in homoclinic tangles with a horseshoe structure~\cite{Smale}. The stochasticity bands originated by the destruction of KAM tori grow in size with the perturbation, and eventually overlap, this giving rise to an scenario of widespread chaos~\cite{Chirikov}.

\vspace{2ex}   

Quantum mechanically, the dynamical characteristics of the vibrational states are usually ascertained by examining the topology and nodal patterns~\cite{StrattHandyMiller} of the corresponding wavefunctions in configuration space. This method works remarkably well for near integrable system even for sizeable values of the perturbation parameter~\cite{Jung}, but gives little clue in the case of strong mixing or ergodic cases~\cite{Shnirelman}. Important in the later case is the phenomenon known as \textit{scarring}~\cite{Heller}, which refers to the localization of quantum density probability along the less unstable POs of a classically chaotic system in a set of measure zero of their eigenfunctions. This concept has been fruitfully~\cite{Fabio2,Fabio3} extended to the case of non-stationary scarred wavefunctions, as shown in Refs.~\cite{Polavieja,Sibert2,Fabio1}. Also, phase space pictures of the quantum states can be constructed by means of quasiprobability density functions, such as the Wigner~\cite{Wigner} or Husimi functions~\cite{Husimi}. The position of the maxima of these functions contains information on the phase space structures relevant in the dynamics of the state~\cite{Haake}, and also the zeros of the Husimi functions do the same job efficiently~\cite{Wisniacki}.

\vspace{2ex}   

In this paper, we present a theoretical study of the dynamical characteristics of the vibrational eigenstates of KCN, which was presented in the past as the first example of early onset of chaos in a bounded molecular systems~\cite{Farantos1}, and was also the subject of classical and quantum studies ~\cite{Tennyson1,Tennyson2} (in an inaccurate potential energy surface; see discussion in Ref.~\cite{Parraga.KCN.PES}). Indeed, the corresponding classical dynamics is very chaotic even at very low values of the excitation energy. For this purpose we use energy levels correlation diagram using $\hbar$ as varying parameter, as previously done in similar molecular systems~\cite{Arranz.LiCN-HCN-HO2.correlation.diagram}. With this method the dynamical characteristics of the different states can be easily unveiled, contrary to what happens in the usual quantum mechanical calculations, in which only the actual value of the Planck's constant, i.e., $\hbar=1$~a.u., is considered. We also establish the frontier for the quantum transition from order to chaos in terms of scarred functions, and its validity for generic  molecular systems with mixed classical dynamics.

\vspace{2ex}   

The organization of this paper is as follows. In the next section, we present the molecular model used and the calculations. First, we describe the Hamiltonian and potential energy surface used. Next, we indicate how the classical dynamics of the molecule is described. And we finish by describing the methods used to compute the quantum vibrational states, the level correlation diagram \textit{vs.}~$\hbar$, and the coupling among states in this representation. in Sec.~\ref{sec:results} we present our results. First, we discuss Poincar\'e surfaces of section at different values of the excitation energy, fully describing the evolution of the KCN phase space with the energy, which show a very early transition to widespread chaos. We continue by presenting the quantum results, paying special attention to the description of the adiabatic correlation diagram for the vibrational energy levels and its characteristics. In particular, we discuss in detail the diabatic states that emerge in it, and present a simple model to explain their origin and dynamical properties. We conclude the section by discussing the existence of a quantum transition from order to chaos, which appears as a frontier of scarred functions. Finally, we present some concluding remarks in Sec.~\ref{sec:conclusion}.

\section{\label{sec:model}Molecular model and calculations}

\subsection{\label{sec:hamiltonian}Hamiltonian molecular model}

The vibrational dynamics of the KCN molecule can be adequately studied~\cite{Parraga.KCN.PES} with a two degrees of freedom, in which the C-N motion is kept frozen at its equilibrium distance due to the existing adiabatic separation from the remaining vibrational modes in the system~\cite{LiCN.3D}.

\vspace{2ex}   

The corresponding Hamiltonian function for the purely vibrational, i.e., without rotation, dynamics of KCN molecule is given in Jacobi coordinates by
\begin{equation}
\label{eq:hamiltonian}
H = \frac{P_R^2}{2\mu_1} + \frac{P_\theta^2}{2} \left( \frac{1}{\mu_1 R^2} 
    + \frac{1}{\mu_2 r_\text{eq}^2} \right) + V(R,\theta),
\end{equation}
where $\mu_1 = m_\text{K}(m_\text{C} + m_\text{N})/(m_\text{K} + m_\text{C} + m_\text{N})$ and $\mu_2 = m_\text{C} m_\text{N}/(m_\text{C} + m_\text{N})$ are reduced masses, being $m_X$ the corresponding atomic masses, $r_\text{eq} = 2.22$ a.u.\ is the frozen \mbox{C-N} equilibrium length, $R$ is the length from the \mbox{C-N} center of mass to the K atom, and $\theta$ is the angle formed by the corresponding $R$ and $r_\text{eq}$ directions, with $\theta=0$ and $\theta=\pi$ rad corresponding to the linear configurations \mbox{K-CN} and \mbox{CN-K}, respectively. $P_R$ and $P_\theta$ are the associated conjugate momenta, and $V(R,\theta)$ is the potential energy function describing the vibrational interactions in the K-CN molecular system.

\vspace{2ex}   

For the potential energy function $V(R,\theta)$, we use the analytic expression, fitted to \textit{ab initio} quantum calculations, of P\'arraga \textit{et al.}~\cite{Parraga.KCN.PES}. This potential energy function has two minima: the deepest absolute minimum at $\theta\approx\pi/2$ rad, corresponding to the triangular molecular configuration $\genfrac{}{}{0pt}{}{\text{K}}{\text{C-N}}$, and the shallow relative minimum at $\theta=0$, associated to the colinear configuration K-C-N. The other colinear configuration, i.e., C-N-K, corresponds to a saddle point rather than to a minimum, which is quite flat in the angular coordinate.

\vspace{2ex}   

As it was shown in Ref.~\cite{Parraga.KCN.PES}, the dynamics of the K-CN molecular system is highly nonlinear, so that broad chaotic regions appear in its classical phase space even at the energy of the quantum ground state $E_1=145$ cm$^{-1}$, and its quantum eigenstates exhibit overlapped Fermi resonances, and a very irregular nodal pattern starting at the second excited state, this having an energy of a mere $E_3=491$ cm$^{-1}$. 

\subsection{\label{sec:classical.calculation}Classical trajectories calculation}

Classical trajectories for KCN are calculated by numerically integrating the Hamilton equations of motion corresponding to Eq.~(\ref{eq:hamiltonian}). Some POs relevant to our work have been obtained by dynamical propagation of symmetry lines~\cite{Arranz.LiCN.bifurcation.diagram}.

\subsection{\label{sec:quantum.calculations}Quantum calculations}

The eigenenergies and eigenfunctions of KCN have been obtained with the Discrete Variable Representation--Distributed Gaussian Basis (DVR-DGB) method of Ba\u ci\'c and Light~\cite{DVR-DGB} applied to the Hamiltonian operator corresponding to (\ref{eq:hamiltonian}). In this way, and using a final basis set of approximately 1000 ray eigenvectors (on average for the different values of $\hbar$) lying in 50 angular rays. In this way, approximately the 300 (on average) low lying eigenfunctions, $\langle R\,\theta\,|n\rangle$ ($n=1,\ldots, 300$), for values $\hbar=\lbrace0.10,0.11,0.12,\ldots,3.00\rbrace$ a.u., with its eigenenergies converged to within 1 cm$^{-1}$ were obtained. Let us indicate that for lower values of $\hbar$, e.g.~0.5 a.u.\ to 0.1 a.u., the number of rays needed to be increased to 120 to maintain accuracy. Let us remark that the results can be strongly dependent on the value of $\hbar$ used in the calculations.

Using the data obtained from this calculation a correlation diagram of eigenenergies or vibrational energy levels is constructed. As varying parameter in this diagram we take Planck physical constant. This may seem an awkward choice at first sight, but this is not the case for the following reasons, which are the justification for our election. When decreasing values of $\hbar$ are considered, the volume of phase space occupied by the quantum states, in the sense of the Weyl's semiclassical prescription~\cite{Weyl}, also decreases. Accordingly, the regular region of phase space in a generic Hamiltonian system, like ours, can ``accommodate'' more states, and then a transition from chaos to order will take place. Alternatively, changes in $\hbar$ can be considered equivalent to the same changes performed in the value of the masses of the atoms forming the molecule. Although nature does not allows a continuum range of isotope masses, it certainly provides us with a limited number of isotopic realtions in the masses involved in the Hamiltonian function (\ref{eq:hamiltonian}). Let us also remark that, in our case and due to numerical convergence problems, the correlation diagram can only be accurately computed for values of $\hbar<0.10$ a.u.

An additional quantity of interest in relation to correlation diagrams are the coupling matrix elements $\langle m|\partial/\partial\hbar|n\rangle$, which determine the interaction (or mixing) between eigenstates when the parameter is varied. In our case, we evaluate these couplings by applying the off-diagonal Hellmann-Feynman theorem ~\cite{Born.adiabatic.theorem,Fernandez.hypervirial.theorems}, in the following way
\begin{equation}
\label{eq:hellmann-feynman.theorem}
\langle m| \frac{\partial}{\partial\hbar} |n\rangle =
\frac{1}{E_n - E_m}\langle m| \frac{\partial\widehat{H}}{\partial\hbar} |n\rangle,
\end{equation}
being the eigenvalues equation $\widehat{H}|m\rangle = E_m|m\rangle$, and $\widehat{H}$ the Hamiltonian operator corresponding to Eq.~(\ref{eq:hamiltonian}). Explicit mathematical expressions for the matrix elements $\langle m|\partial\widehat{H}/\partial\hbar|n\rangle$ in Eq.~(\ref{eq:hellmann-feynman.theorem}) for the specific case of Ba\u ci\'c and Light's DVR-DGB wavefunctions can be found in Ref.~\cite{Arranz.LiCN.correlation.diagram}.

\section{\label{sec:results}Results and discussion}

\subsection{\label{sec:classical.results}Classical dynamics: Poincar\'e surface of section}

The classical dynamics of KCN can be studied by calculating composite Poincar\'e surfaces of section (PSOS) for a selection of representative values of the excitation energy in the way prescribed in Ref.~\cite{Ezra}.

Some results can be found in Ref.~\cite{Parraga.KCN.PES}, where the route to chaos is carefully discussed. First, the KCN regular dynamics at very low energies, i.e., below 65~cm$^{-1}$, is organized around the triangular configuration (central 1:1 resonance) and another 1:2 (asymmetric motion) resonance. Above that energy, chaos quickly sets in, and by $E=290$~cm$^{-1}$ very few traces of order remain. Later a very interesting phenomenon occurs at $1200\le E \le 5400$~cm$^{-1}$ since an important region of order above the K-NC saddle appear in the middle of a sea of chaos~\cite{JPCA}. Even later, above 7500~cm$^{-1}$ small islands of regularity appear corresponding to the hinge motions around the K-CN and K-NC linear configurations described below.

It will also be discussed in Sec.~\ref{sec:diabatic.correlation.diagram}, that the stable structures at the colinear configurations, and the stable structures corresponding to hinge POs are both relevant for the emergence of diabatic states in our correlation diagrams of eigenenergies.

\subsection{\label{sec:quantum.results}Quantum results}

\subsubsection{\label{sec:adiabatic.correlation.diagram}Adiabatic correlation diagram for the\\ KCN vibrational energy levels}

\begin{figure*}
\includegraphics{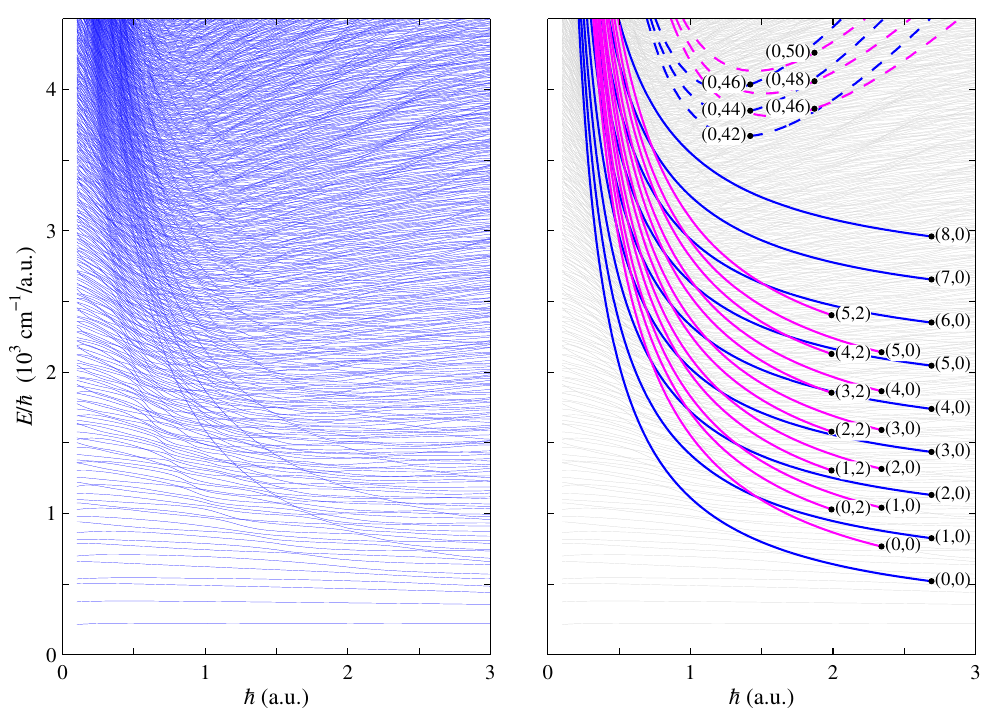}
\caption{\label{fig:correlation.diagram}(Left) Correlation diagram for the KCN adiabatic vibrational energy levels \textit{vs.}\ $\hbar$. The emergence of diabatic states formed by interaction between eigenstates at the avoided crossings is observed. Consider, for example, the hyperbolic ``curves'' specially visible in the lower left part of the figure, and also the (approximately) straight lines at the top right of it (see text for details). (Right) Correlation diagram for the KCN diabatic vibrational energy levels, obtained from Eq.~(\ref{eq:HO.model}) (solid lines), and Eq.~(\ref{eq:HR.model}) (dashed lines), \textit{vs.}\ $\hbar$. Magenta/lighter and blue/darker curves correspond to states around the K-CN ($\theta=0$) and K-NC ($\theta=\pi$ rad.) linear configurations, respectively. All curves are labeled by the corresponding diabatic quantum numbers $(n_1,n_2)$. In both graphs energy in the vertical axis has been scaled with $\hbar$ to obtain clearer plots which would appear otherwise too crowded near the the origin(see text for details).}
\end{figure*}
Figure~\ref{fig:correlation.diagram}~(left) shows the correlation diagram of KCN vibrational energy levels \textit{vs.}\ $\hbar$ (taken as a variable parameter) computed with the DVR-DGB program of Ba\u ci\'c and Light described in Sec.~\ref{sec:quantum.calculations}. The eigenenergies have been scaled with respect to $\hbar$ in the vertical axis in order to obtain a clearer graphical representation, which would otherwise be too crowded near the origin. Notice that this particular scaling transformation renders horizontal lines for completely harmonic energy levels, that would otherwise accumulate as $\hbar\rightarrow0$. Several comments are in order.

First, the curves in the correlation diagram on the left plot of Fig.~\ref{fig:correlation.diagram} present a very complicated ``spaghetti-like'' structure. Despite this fact, two relevant features are clearly observed in it. For one thing, the eigencurves avoid crossing at many places. This behavior is a consequence of the well known von Neuman non-crossing rule~\cite{non.crossing.rule}, and in this sense these states are said to be \textit{adibatic} with respect to changes in the parameter. Also, it is seen the existence of a widespread important level repulsion at the avoided crossings (ACs), except in the left bottom part of the correlation diagram. According to the Bohigas-Giannoni-Schmit conjecture of random matrix theory~\cite{Metha.random.matrix.theory}, this is a consequence of the underlying classical chaotic behavior of the system, and it can be considered accordingly as an indication of the existence of quantum chaos~\cite{Haake,Marcus}. Moreover, the region of regularity existing near $\hbar\rightarrow0$, i.e., in the semiclassical limit, is very small in this case, contrary to what has been previously observed in calculations for other similar molecular systems; see, for example, Ref.~\cite{Arranz.LiCN-HCN-HO2.correlation.diagram}. We will return to this point, which in some sense can be understood as a quantum order-chaos transition taking place in the correlation diagram, later in Sec.~\ref{sec:order-chaos.transition}.

Second, when the correlation diagram of the left plot of Fig.~\ref{fig:correlation.diagram} is visually inspected from a distance, or somehow blurring the fine details, a series of hyperbolic shaped ``continuous curves'' seem to exist in the bottom left part of the plot, as well as other group of ``continuous lines'' at the top right part, which are asymptotically linear. However, when closely examined these apparent ``continuous curves'' are seen to consist of series of narrow ACs among the different adiabatic states. Nevertheless, Schmidt in 1969~\cite{Schmidt.diabatic} showed that the states corresponding to the ``continuous curves'' can be constructed as linear combinations of the adiabatic states, in such a way that their interactions~\cite{Landau.Zener} vanish, so that the new states can cross when they approach. He coined the term \textit{diabatic} for these states, and they have the property of maintaining some sort of character~\cite{Borondo1} as the parameter of the correlation diagram changes. This character can usually be defined in terms of conserved physical characteristic magnitudes and their corresponding quantum numbers~\cite{Borondo2}, or actions if the classical counterpart is considered.
\begin{table}[t!]
\caption{\label{tab:models.parameters}Numerical values for the parameters entering in Eqs.~(\ref{eq:HO.model}) and (\ref{eq:HR.model}) for the harmonic oscillator (HO) and hindered rotor (HR) models at the K-CN and K-NC linear configurations.}
\begin{ruledtabular}
\begin{tabular}{cccccc}
      & $\theta$ & $\epsilon_0$  & $\omega_1$             & $\omega_2$
      & $B$                          \\
Model & (rad)    & (cm$\!^{-1}$) & (cm$\!^{-1\!}\!/$a.u.) & (cm$\!^{-1\!}\!/$a.u.)
      & (cm$\!^{-1\!}\!/$a.u.$\!^2$) \\
\hline
HO K-CN &     0 &           1376 & 275 & 80 &      \\
HR K-CN &     0 &           2900 & 275 &    & 0.55 \\
HO K-NC & $\pi$ & \phantom{0}940 & 305 & 37 &      \\
HR K-NC & $\pi$ &           2400 & 305 &    & 0.73 \\
\end{tabular}
\end{ruledtabular}
\end{table}
\begin{table}[t!]
\caption{\label{tab:eigenstates.parameters}Characteristics, parameters, and adiabatic and diabatic quantum numbers identifying the eigenstates in Figs.~\ref{fig:wf.harmonic.minimum}, \ref{fig:wf.harmonic.saddle}, and \ref{fig:wf.hinge}, corresponding to emerging diabatic harmonic oscillator and hinge states around both linear K-CN and K-NC linear configurations in the correlation diagram of Fig.~\ref{fig:correlation.diagram}.}
\begin{ruledtabular}
\begin{tabular}{lcccc}
\multicolumn{1}{c}
{Characteristics} & $\hbar$ (a.u.) & $E$ (cm$^{-1}$) & $n$ & $(n_1,n_2)$ \\
\hline
Harmonic on K-CN & 0.621  & 1509 & \phantom{0}92 & $(0,0)$     \\
                 & 0.525  & 1556 &           140 & $(0,2)$     \\
                 & 0.628  & 1687 &           120 & $(1,0)$     \\
                 & 0.590  & 1749 &           148 & $(1,2)$     \\
                 & 0.606  & 1845 &           161 & $(2,0)$     \\
                 & 0.644  & 1971 &           166 & $(2,2)$     \\
                 & 0.658  & 2069 &           180 & $(3,0)$     \\
                 & 0.632  & 2144 &           210 & $(3,2)$     \\
                 & 0.836  & 2485 &           168 & $(4,0)$     \\
                 & 0.825  & 2612 &           192 & $(4,2)$     \\
                 & 0.664  & 2439 &           255 & $(5,0)$     \\
                 & 0.680  & 2591 &           278 & $(5,2)$     \\
\hline
Harmonic on K-NC & 0.366  & 1006 & \phantom{0}99 & $(0,0)$     \\
                 & 0.420  & 1127 &           100 & $(1,0)$     \\
                 & 0.528  & 1347 & \phantom{0}97 & $(2,0)$     \\
                 & 0.432  & 1406 &           160 & $(3,0)$     \\
                 & 1.160  & 2557 & \phantom{0}93 & $(4,0)$     \\
                 & 1.060  & 2719 &           127 & $(5,0)$     \\
                 & 0.798  & 2524 &           191 & $(6,0)$     \\
                 & 0.800  & 2756 &           230 & $(7,0)$     \\
                 & 0.930  & 3311 &           247 & $(8,0)$     \\
\hline
Hinge on K-CN    & 2.200  & 8854 &           277 & $(0,46)$    \\
                 & 2.100  & 8790 &           300 & $(0,48)$    \\
                 & 1.980  & 8570 &           322 & $(0,50)$    \\
\hline
Hinge on K-NC    & 2.160  & 8750 &           282 & $(0,42)$    \\
                 & 2.060  & 8720 &           308 & $(0,44)$    \\
                 & 1.910  & 8330 &           331 & $(0,46)$    \\
\end{tabular}
\end{ruledtabular}
\end{table}

Third, there are two kinds of such ``continuous curves'' in the correlation diagram in Fig.~\ref{fig:correlation.diagram} (left). Some of them are more sharply defined since the energy separation at the ACs are very small (actually not appreciable in the scale of the figure), while the others are more diffuse, this group corresponding to larger energy separations at the ACs. This effect is most clearly seen in the first group of hyperbolic ``continuous curves'' than in the second, where the correlation diagram is much more crowded.

Diabatic states provides a great help in unveiling the physical characteristics of the involved adiabatic eigenstates, which will be otherwise get lost due to the presence of numerous overlapping interactions. Furthermore, it has also been shown that these diabatic structures constitute the frontier in the energy level correlation diagrams where the semiclassical phenomena of scarring~\cite{Heller} first appears in energy~\cite{Arranz.LiCN.bifurcation.diagram,Arranz.LiCN.correlation.diagram,Arranz.LiCN.scars.edge}.

\subsubsection{\label{sec:diabatic.correlation.diagram}Diabatic vibrational states for KCN and correlation diagram}

In order to construct the diabatic vibrational states for KCN that are observed emerging in the correlation diagram of Fig.~\ref{fig:correlation.diagram} (left), we will use here the following model/approximation for the corresponding energy levels. For simplicity, we will only consider vibrational structures around the two linear configurations, K-CN ($\theta=0$) and K-NC ($\theta=\pi$ rad). We will also assume that in both of them the contributions from each mode, i.e., K-CN/K-NC stretching corresponding to the $R$ coordinate, and K-C-N bending corresponding to $\theta$, are separable.

\begin{figure*}[t!]
\includegraphics{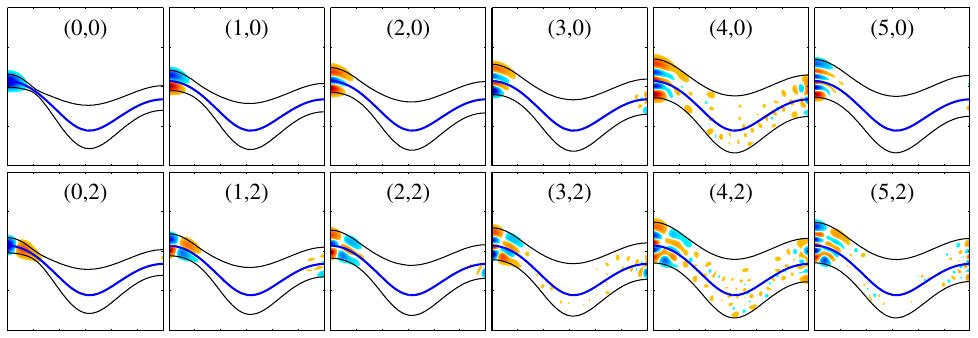}
\caption{\label{fig:wf.harmonic.minimum}Some adiabatic eigenstates of KCN that can be assigned to harmonic oscillator states on the K-CN linear configuration at $\theta=0$. The probability density is represented in a color/gray scale. The minimum energy path and the corresponding eigenenergy contour have also been represented as blue thick line and black thin line, respectively. Appropriate diabatic quantum numbers $(n_1,n_2)$ [as in Eq.~({\ref{eq:HO.model}})] are given in each plot. The values of $\hbar$ at which each of these states have been calculated is reported in Table~\ref{tab:eigenstates.parameters}. The horizontal and vertical axis span the ranges $[0,\pi]$ rad and $[4,8]$ a.u., respectively.}
\end{figure*}
These two contributions are then computed using the harmonic oscillator (HO) model with a quadratic potential energy function, and the hindered rotor (HR) model~\cite{Barker.hindered.rotor} corresponding to a sinusoidal potential function, respectively. For the HO, the energy levels are given by $\epsilon_m=V_\text{min}+\hbar\omega(m+1/2)$, where $V_\text{min}$ is the minimum of the potential energy function, and $\omega$ and $m$ correspond to the oscillator frequency and the quantum number, respectively. For the HR we have two possibilities. At low energy, i.e., well below the sinusoidal energy barrier $V_\text{max}$, the energy levels are also given in a first approximation by a harmonic oscillator expression~\cite{Barker.hindered.rotor}. Accordingly, we can consider that, at low energies, the KCN diabatic energies, $\epsilon$, are given by the expression
\begin{equation}
\label{eq:HO.model}
\epsilon_{n_1 n_2} = \epsilon_0 
+ \hbar \left[ \omega_1 \left( n_1 + \frac{1}{2} \right) 
+ \omega_2 \left( n_2 + \frac{1}{2} \right) \right],
\end{equation}
where $\epsilon_0$ is the absolute minimum of the potential function (at the triangular configuration), and subscripts 1 and 2 refer to stretching and bending motions, respectively~\cite{footnote1}. On the other hand, at high energies, i.e., well above the bending energy barrier $V_\text{max}$, the energies can be approximated by the free rotor expression displaced by half energy barrier~\cite{Barker.hindered.rotor}, i.e., $\epsilon_m=V_\text{max}/2+\hbar^2 B m^2$, being $B$ (defined $\hbar$ free) the corresponding rotational constant. Accordingly, at high energies, the KCN diabatic energy levels are given by
\begin{equation}
\label{eq:HR.model}
\epsilon_{n_1 n_2} = \epsilon_0 
+ \hbar\omega_1 \left( n_1 + \frac{1}{2} \right) 
+ \hbar^2 B n_2^2,
\end{equation}
where the quadratic term (both in $\hbar$ and the quantum number) corresponds to the bending motion, the linear term to the stretching motion, and the independent term $\epsilon_0$ gathers possible minimum energy from stretching and half energy barrier from bending. Notice that in Eqs.~(\ref{eq:HO.model}) and (\ref{eq:HR.model}) $\varepsilon_0, \omega_1, \omega_2$, and $B$ have different values for HO, HR, and the two possible linear configurations, K-CN and K-NC. Suitable values for these parameters are given in Table~\ref{tab:models.parameters}. They have been obtained from the minimum potential energy path and normal mode frequencies of KCN given in Ref.~\cite{Parraga.KCN.PES}. In the case of the saddle point ($\theta=\pi$ rad.), the real modulus of the bending frequency is taken instead of the original purely imaginary one $i\omega_2$.

The corresponding results for the correlation diagram of KCN diabatic vibrational levels obtained from Eqs.~(\ref{eq:HO.model}) (solid lines) and (\ref{eq:HR.model}) (dashed lines) both at the K-CN ($\theta=0$) (magenta/lighter lines), and K-NC ($\theta=\pi$ rad.) (blue/darker lines) linear configurations, are shown in the right plot of Fig.~\ref{fig:correlation.diagram}. As can be seen, the diabatic level curves here appear segregated into two different groups. The first one (solid magenta and solid blue lines) presents a hyperbolic dependence with $\hbar$, which is due to the existence of the non-zero term $\epsilon_0$ in Eq.~(\ref{eq:HO.model}). A second group (dashed magenta and blue lines) appears at the top right part of the figure, with a distorted parabolic form, which is asymptotically linear.

Let us start by discussing the first group. When the corresponding (approximate) diabatic curves are compared with those in the adiabatic correlation diagram on the left plot of Fig.~\ref{fig:correlation.diagram}, one observes that the solid lines, obtained from Eq.~(\ref{eq:HO.model}), nicely match the hyperbolic emerging diabatic states in the adiabatic diagram. This provides a quantitative confirmation of the existence and origin of these diabatic states in KCN. This comparison can be extended further, to the more diffuse diabatic states, some of which are clearly visible at the leftmost part of the adiabatic correlation diagram in Fig.~\ref{fig:correlation.diagram}, can be assigned in our model as corresponding to states located over the K-NC linear configuration (blue/darker lines), while those which are more sharply defined are similar states but located on the K-CN linear configuration (magenta/lighter lines) of the potential.
\begin{figure}[t!]
\includegraphics{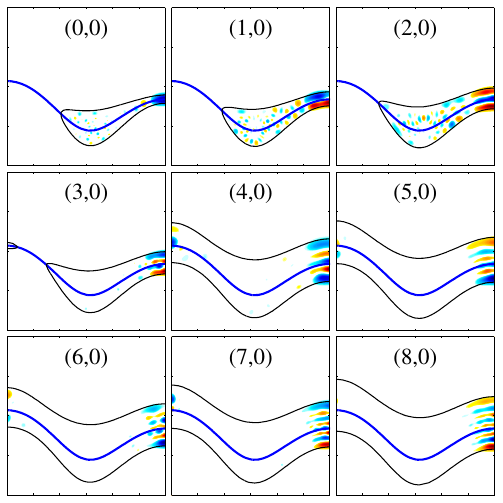}
\caption{\label{fig:wf.harmonic.saddle}Same as Fig.~\ref{fig:wf.harmonic.minimum} for harmonic oscillator states on the K-NC colinear configuration at $\theta=\pi$ rad. The values of $\hbar$ at which each of these states have been calculated is reported in Table~\ref{tab:eigenstates.parameters}.}
\end{figure}

To make this comparison and the corresponding state assignment more precise, we have selected some representative states of both classes, i.e., diffuse and sharp diabatic states. The parameters defining their position in the correlation diagram of Fig.~\ref{fig:correlation.diagram} and their dynamical characteristics are given in the first two groups of Table~\ref{tab:eigenstates.parameters}, along with the adiabatic eigenstate number, and the corresponding diabatic quantum numbers in Eq.~(\ref{eq:HO.model}). The corresponding wavefunctions are presented in Fig.~\ref{fig:wf.harmonic.minimum} and Fig.~\ref{fig:wf.harmonic.saddle} for eigenstates on the K-CN and K-NC linear configurations, respectively. As can be seen, in these two figures the selected eigenfunctions clearly have a harmonic oscillator character, with a nodal structure corresponding to the quantum numbers, predicted in our discussion above.

One final point is worth discussing here. It concerns the relationship existing between the diffuse/sharp character of the diabatic states located, respectively, on the linear K-CN minimum ($\theta=0$) and the linear K-NC saddle ($\theta=\pi$ rad.) of the potential energy surface, and how much their densities extend along the bending coordinate. Indeed, as can inferred from a careful observation of the wavefunctions in Figs.~\ref{fig:wf.harmonic.minimum} and \ref{fig:wf.harmonic.saddle}, the latter, i.e., states located over the saddle, only appear excited along the stretching coordinate, with $n_2=0$, and they give rise to diffuse diabatic states in the correlation diagram. On the contrary, the former, i.e., states localized on the linear minimum, are eventually excited, yet extended, along the bending coordinate, i.e., with $n_2 \ne 0$, and they give rise to the sharp diabatic states. Obviously, this effect can be correlated with the stability of the corresponding phase space region investigated in Refs.~\cite{Parraga.KCN.PES,JPCA}. In this respect, recall that large region of regularity were found located around both the K-CN and K-NC linear configurations. The quantization of tori in this region should lead to the definition of diffuse and  sharp diabatic states.

\begin{figure}[t!]
\includegraphics{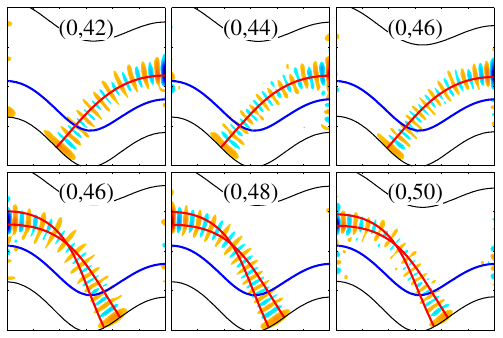}
\caption{\label{fig:wf.hinge}Same as Fig.~\ref{fig:wf.harmonic.minimum} for hinge eigenstates around K-NC (top) and K-CN (bottom) linear configurations. The values of $\hbar$ at which each of these states have been calculated is reported in Table~\ref{tab:eigenstates.parameters}. The involved stable periodic orbits have been represented as red/lighter thick lines.}
\end{figure}
Let us concentrate now in the discussion of the second group of ``continuous curves'' which appear at the top right of the correlation diagram  in Fig.~\ref{fig:correlation.diagram} (left). To obtain the corresponding diabatic states, we apply the HR model of Eq.~(\ref{eq:HR.model}), and the results are shown in dashed (magenta and blue) lines in the correlation diagram on the right plot of Fig.~\ref{fig:correlation.diagram}. Observe that in the $E/\hbar$ \textit{vs.}\ $\hbar$ representation, the HR expression leads to rational function curves, with hyperbolic behavior at low values of $\hbar$ (and a vertical asymptote at $\hbar = 0$, due to the independent term), and a linear dependence for high $\hbar$ values (and a oblique asymptote $E/\hbar=\omega_1/2+\hbar B n_2^2$, due to the quadratic term). By comparing the diabatic HR curves and the adiabatic ones in the right and left plots of Fig.~\ref{fig:correlation.diagram}, it is observed that diabatic states emerging as mainly straight curves at the top right of the figure match with the HR states for high $\hbar$ values (oblique asymptote).

Again, and similarly to what we did in the HO case, we can make this comparison more precise, by selecting some curves in the adiabatic correlation diagram at the top right of the left plot in Fig.~\ref{fig:correlation.diagram} which show an asymptotic linear behavior. The identifying parameters corresponding to those curves are reported in the two last groups of Table~\ref{tab:eigenstates.parameters}, and the corresponding eigenfunctions are given in Fig.~\ref{fig:wf.hinge}. Surprisingly enough, the selected eigenstates correspond to the so called \textit{hinge} states, i.e., states with the probability density highly localized over stable periodic orbits with hinge motion (also depicted in the figure), rather than o rotor-like states, as should be expected a priori. Moreover, those states appear with two different values of the asymptotic slope, corresponding to hinge states around the two possible linear configurations of the molecule (see values in Table~\ref{tab:models.parameters}). Actually, true rotor-like states, corresponding to mainly straight line diabatic states in the correlation diagram $E/\hbar$ \textit{vs.}\ $\hbar$, had already been described in the Li-CN molecular system~\cite{Arranz.LiCN.correlation.diagram}. This makes of this case, i.e., the hinge states behavior described here, especially interesting. In first approximation, hinge states can be considered as belonging the system defined by the angular coordinate $\theta$ subjected to the energy potential given by the energy profile along the corresponding hinge periodic orbit. Under this approximation, the motion around each linear configuration ($\theta=0$ or $\theta=\pi$ rad), that is, in the range of the coordinate $\theta$ bounded by the intersection between hinge periodic orbit and minimum energy path, approximately reproduces a sinusoidal hindered rotor. It is interesting to note that, in this case, the corresponding half-energy barrier $V_\text{max}/2$ of the hindered rotor model is similar (albeit slightly smaller) to the fitting parameter $\varepsilon_0$.

\subsubsection{\label{sec:order-chaos.transition}Quantum transition from order to chaos in the correlation diagram of KCN}

One final point worth discussing in this work is in connection with the (apparent) non-existence of a clear regular region in the lower left region of the correlation diagram of KCN, as it is the case in other similar floppy molecules, such as HCN, LiCN, or HO$_2$~\cite{Arranz.LiCN-HCN-HO2.correlation.diagram,Arranz.LiCN.correlation.diagram}. Actually, in the semiclassical limit $\hbar \rightarrow 0$ the quantum states ``occupy'' less and less phase space volume in the sense discussed by Weyl~\cite{Weyl}. Since at the same time the classical phase space remains unchanged, a chaos-order transition is expected as the quantum states get smaller and then migrate to the regular region of phase space. As previously found, this transition takes place as the frontier for the appearance of scaring~\cite{Heller} is crossed in the correlation diagram~\cite{Arranz.LiCN-HCN-HO2.correlation.diagram,Arranz.LiCN.correlation.diagram,Arranz.LiCN.scars.edge}.

\begin{figure}[t!]
\includegraphics{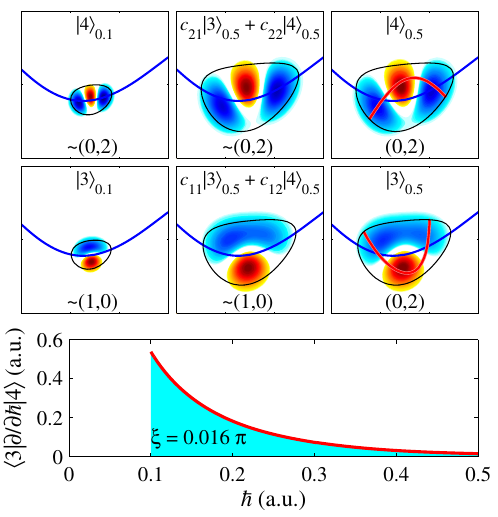}
\caption{\label{fig:wf.scars.frontier_3-4}(Bottom) Coupling between eigenstates $|3\rangle$ and $|4\rangle$ in the range $\hbar\in[0.1,0.5]$ a.u. The value of the mixing angle $\xi$, corresponding to the area under the curve, is also indicated. In the upper panels, the wavefunctions corresponding to the eigenstates $|m\rangle_\hbar$ at $\hbar=0.1$ a.u.\ (left column), $\hbar=0.5$ a.u.\ (right column), and the unmixed states obtained from Eqs.~(\ref{eq:orthogonal.transformation}) and (\ref{eq:mixing.angle}) (middle column) are represented in color/gray scale. The minimum energy path and the potential contour corresponding to the eigenenergies (expectation energy $\langle\widehat{H}\rangle$, in the unmixed cases) have also been represented as blue thick line and black thin line, respectively. The involved 1:2 periodic orbits have been represented as red/lighter thick lines. Appropriate quantum numbers $(n_1,n_2)$ are also indicated. Axes are the same as in Fig.~\ref{fig:wf.regular.frontier}.}
\end{figure}

\begin{figure}[t!]
\includegraphics{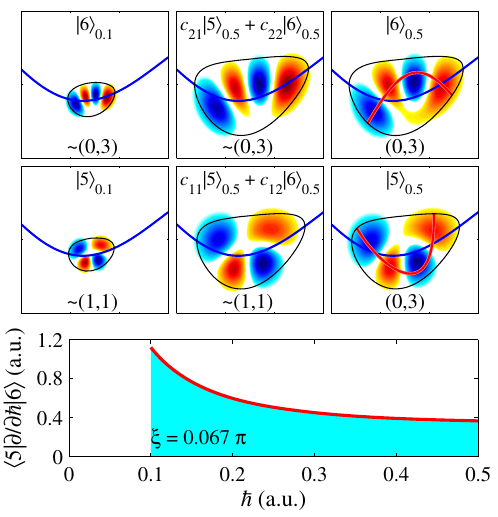}
\caption{\label{fig:wf.scars.frontier_5-6}Same as in Fig.~\ref{fig:wf.scars.frontier_3-4} for eigenstates $|5\rangle$ and $|6\rangle$.}
\end{figure}

In order to examine this problem in detail, we will perform a quantitative analysis by focusing our attention in the lowest lying eigenstates presenting a non-regular $(n_R,n_\theta)$ nodal pattern (in the sense discussed in Ref.~\cite{StrattHandyMiller}) near the region at $\hbar=0.1$ a.u. This let aside the ground state $|1\rangle$, which is mostly Gaussian, and first excited $|2\rangle$ state, which presents a mostly $(n_R,n_\theta)=(0,1)$ regular wavefunction~\cite{Parraga.KCN.PES}), and then we concentrate in the next eigenstates $|3\rangle$, $|4\rangle$, $|5\rangle$, and $|6\rangle$. As shown in the two upper right panels of Figs.~\ref{fig:wf.scars.frontier_3-4} and~\ref{fig:wf.scars.frontier_5-6} the corresponding wavefunctions exhibit (for $\hbar=0.5$ a.u.) non-regular $(n_R,n_\theta)$ nodal patterns. To unveil the vibrational structure underlying that non-regular topology we will assume for simplicity that it arises from the interaction between pairs of eigenstates, i.e., $|3\rangle$ and $|4\rangle$, and $|5\rangle$ and $|6\rangle$, due to quantum resonances between vibrational modes $\Delta n_R:\Delta n_\theta$~\cite{McHale}, as it is frequently the case~\cite{Arranz.LiCN.correlation.diagram}. The corresponding diabatic states can be obtained  by means of the following (orthogonal~\cite{Oregi.Henon-Heiles.Husimi.zeros}) transformation
\begin{equation}
\label{eq:orthogonal.transformation}
\begin{pmatrix}
|\chi_i\rangle \\
|\chi_j\rangle
\end{pmatrix}
=
\begin{pmatrix}
\ \ \cos\xi & \sin\xi \\
   -\sin\xi & \cos\xi
\end{pmatrix}
\begin{pmatrix}
|m\rangle \\
|n\rangle
\end{pmatrix}
,
\end{equation}
where the mixing angle $0\le\xi\le \pi/2$ is determined by the differential equation
\begin{equation}
\label{eq:mixing.angle}
\frac{\text{d}\xi}{\text{d}\hbar} =
\langle m| \frac{\partial}{\partial\hbar} |n\rangle,
\end{equation}
so that $\xi$ is given by the area under the coupling curve $\langle m|\partial/\partial\hbar|n\rangle$ as a function of $\hbar$.

In the bottom panel of Fig.~\ref{fig:wf.scars.frontier_3-4} we present the coupling curve between eigenstates $|3\rangle$ and $|4\rangle$ in the range from $\hbar=0.1$ a.u.\ to $\hbar=0.5$ a.u.\ (recall that this range cannot be extended to lower values of $\hbar$ for numerical problems). It should have a bell-shaped form, but in our calculation only the right tail of it is visible, and the region around the maximum and the left tail, necessarily to perform a complete analysis of the associated diabatic states, are not available. We also present in the top panels of the figure the involved eigenfunctions for $\hbar=0.1$ a.u.\ (top left panels) and $\hbar=0.5$ a.u.\ (top right panels), where it can be observed that they both present a non-regular probability density pattern, which is however localized along the two stable and unstable 1:2 classical POs (also plotted superimposed in the panels) with quantum numbers $(0,2)$ along each PO path. These POs correspond to the two main 1:2 classical resonances discussed in Sec.~\ref{sec:classical.results}. More specifically, they are clearly recognizable in the PSOS for $E=250$ cm$^{-1}$ in Ref.~\cite{Parraga.KCN.PES}. Nevertheless, in the other extreme, i.e., $\hbar=0.1$ a.u.\ (left panels), the eigenstates appear a little bit more regular, although not very much, somewhat corresponding to states $(n_R,n_\theta)=(1,0)$ and $(0,2)$, respectively. Notice that this quantum numbers assignment leads to a quantum resonance order of $|\Delta n_1|$:$|\Delta n_2|=1:2$, in correspondence with the $\nu_R:\nu_\theta=1:2$ classical resonance, according to the properties shown by states like those that we are considering in the scars frontier of scars as described in Refs.~\cite{Arranz.LiCN-HCN-HO2.correlation.diagram,Arranz.LiCN.correlation.diagram,Arranz.LiCN.scars.edge}. Similar results are obtain, i.e., see top middle panels in Fig.~\ref{fig:wf.scars.frontier_3-4}, when trying to unmix the eigenstates at $\hbar=0.5$ a.u., by applying the inverse orthogonal transformation to (\ref{eq:orthogonal.transformation}) with $\xi=0.016\pi$. Despite this unconclusive result, it seem reasonable to assume that the coupling curve in Fig.~\ref{fig:wf.scars.frontier_3-4} is the tail of the lowest lying 1:2 quantum resonance from the frontier of scars in the KCN molecule.

\begin{table}[b!]
\caption{\label{tab:mixing.eigenstates.parameters}Parameters identifying the eigenstates depicted in Figs.~\ref{fig:wf.scars.frontier_3-4} and \ref{fig:wf.scars.frontier_5-6}, corresponding to the transition from order to chaos. The symbol ``$\sim$'' has been used with the meaning of ``approximately''.}
\begin{ruledtabular}
\begin{tabular}{ccccc}
$\hbar$ (a.u.) & $E$(cm$^{-1}$) & $n$ & $(n_1,n_2)$ & Characteristics \\
\hline
0.1 & \phantom{0}50.19 & 3 & $\sim$$(0,2)$\phantom{$\sim$} & $\sim$regular   \\
0.1 & \phantom{0}53.57 & 4 & $\sim$$(1,0)$\phantom{$\sim$} & $\sim$regular   \\
0.1 & \phantom{0}65.80 & 5 & $\sim$$(0,3)$\phantom{$\sim$} & $\sim$regular   \\
0.1 & \phantom{0}69.88 & 6 & $\sim$$(1,1)$\phantom{$\sim$} & $\sim$regular   \\
0.5 & 250.57           & 3 &       $(0,2)$                 &       localized \\
0.5 & 272.33           & 4 &       $(0,2)$                 &       scar      \\
0.5 & 322.64           & 5 &       $(0,3)$                 &       localized \\
0.5 & 353.99           & 6 &       $(0,3)$                 &       scar      \\
\end{tabular}
\end{ruledtabular}
\end{table}
Similar results and conclusions are obtained for the eigenstates $|4\rangle$ and $|5\rangle$, as shown in Fig.~\ref{fig:wf.scars.frontier_5-6}, but in this case the transition to regularity as $\hbar$ decreases is more clear (although still not totally conclusive). Here, the non regular eigenstates $|5\rangle$ and $|6\rangle$ at $\hbar=0.5$ a.u.\ (right panels) are localized over the stable and unstable, respectively, 1:2 classical resonances with quantum numbers $(0,3)$ along each periodic orbit path, and for $\hbar=0.1$ a.u.\ (left panels) tend to regular states with approximate quantum numbers $(n_R,n_\theta)=(1,1)$ and $(0,3)$, leading to the order of resonance $|\Delta n_1|$:$|\Delta n_2|=1$:2 for the quantum resonance. So that, we can assume that this case corresponds to the second lowest lying 1:2 quantum resonance states from the frontier of scars. The parameters and characteristics identifying both pairs of eigenstates discussed above have been summarized in Table~\ref{tab:mixing.eigenstates.parameters} for informative purposes.

\begin{figure}[t!]
\includegraphics{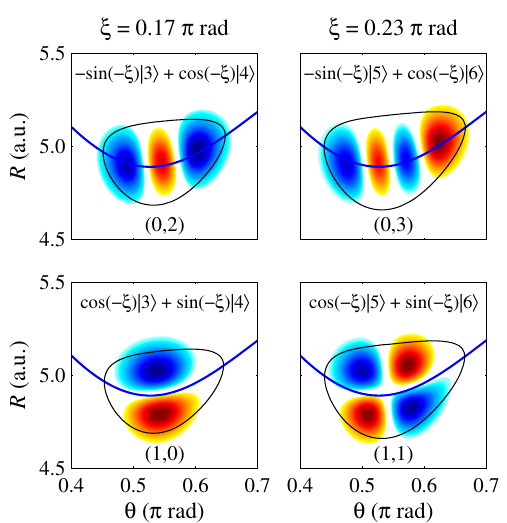}
\caption{\label{fig:wf.regular.frontier}Regular wavefunctions corresponding to the complete unmixing of the eigenstates $|3\rangle$ and $|4\rangle$ (left column), and the eigenstates $|5\rangle$ and $|6\rangle$ (left column), both at $\hbar=0.5$ a.u., represented in color/gray scale. The minimum energy path and the corresponding expectation energy $\langle\widehat{H}\rangle$ contour have also been represented as blue thick line and black thin line, respectively. Appropriate quantum numbers $(n_R,n_\theta)$ are also given.}
\end{figure}
To further check our assumption that the coupling curves in Figs.~\ref{fig:wf.scars.frontier_3-4} and \ref{fig:wf.scars.frontier_5-6} are the tails of the lowest lying 1:2 quantum resonance from the frontier of scars in the KCN molecule, we can extend the unmixing beyond $\hbar=0.1$ a.u.\ by heuristically increasing the mixing angle value $\xi$ until obtaining a regular nodal pattern. The results, i.e., the corresponding diabatic (regular) states, are presented in Fig.~\ref{fig:wf.regular.frontier}, and they prove that the eigenstates localized on the stable and unstable (scarred states) 1:2 classical resonances (POs) at $\hbar=0.5$ a.u.\ are generated by the corresponding 1:2 quantum resonances, that constitute the frontier of scars between the order region and the mixed-chaos region in molecular systems.

\section{\label{sec:conclusion}Concluding remarks}

An analysis of the highly nonlinear dynamics of the KCN molecular system has been carried out in the framework of the quantum manifestations of chaos. For this purpose, a correlation diagram of eigenenergies \textit{vs.}\ the Planck constant, taken as a variable parameter, has been used as the analyzing tool.

\vspace{0.1ex}   

The classical dynamics of this system, as shown by means of composite Poincar\'e surfaces of section at increasing energies, corresponds essentially to mixed-chaos behavior even at very low values of the excitation energy (compared to the fundamental eigenenergy at $\hbar=1$ a.u.). Accordingly, the phase space appears mostly as a sea of chaos with some regular emerging structures, i.e., Kolmogorov-Arnold-Moser tori. As a direct consequence, the quantum correlation diagram shows widely energy level repulsion, corresponding to this classical sea of chaos. Additionally, the correlation diagram shows groups of emerging diabatic states, which correspond to the classical regular tori embedded in the sea of chaos. These regular states, that have been characterized through its wavefunctions and the quantum numbers assignment, are easily identifiable in the correlation diagram, although they can be found difficult to observe at fixed values of the Planck constant (typically $\hbar=1$ a.u.).

\vspace{0.1ex}   

Moreover, we could establish the existence of the frontier, marked by the appearance of scarring~\cite{Heller}, between order and chaos similarly to what happens in other similar molecular systems. In this case, regularity is squeezed into the region of very low $\hbar$ values in the correlation diagram. However, by analyzing the tails of the corresponding couplings, the two lowest lying examples from the frontier of scars have been identified. Thereby, it has been inferred that in this system the frontier of scars arise due to a 1:2 quantum resonance, and results in pairs of eigenstates localized on the corresponding 1:2 stable and unstable (scarred state) classical resonant orbits.

\begin{acknowledgments}
We thank one of our referees for helping us to improve the Introduction with the image of the $\hbar$-microscope in phase space.

This work has been partially supported by the Spanish Ministry of Science, Innovation and Universities, Gobierno de Espa\~na, under Contracts No.\ PGC2018-093854-BI00, and ICMAT Severo Ochoa SEV-2015-0554, and from the People Programme (Marie Curie Actions) of the European Union's Horizon 2020 Research and Innovation Program under Grant No. 734557.
\end{acknowledgments}


\begin{thebibliography}{00}
\bibitem{IachelloLevine}
        F. Iachello and R. D. Levine,
        \textit{Algebraic Theory of Molecules}
        (Oxford University Press, Cambridge, 1995).
\bibitem{KellmanRice}
        M. E. Kellman, F. Amar, and R. S. Berry,
        \textit{Correlation diagrams for Rigid and Nonrigid Three-Body Systems},
        J. Chem. Phys. \textbf{73}, 2387 (1980).
\bibitem{Bunker}
        P. R. Bunker and D. J. Howe,
        \textit{Quantitative energy level correlations for linear, bent and internally rotating triatomic molecules},
        J. Mol. Spectros. \textbf{83}, 288 (1980).
\bibitem{McQuarrie}
        D. A. McQuarrie,
        \textit{Quantum Chemistry}
        (Viva Books, New Delhi, 2016).
\bibitem{Rice}
        D. B. McDonald and S. A. Rice,
        \textit{A correlation diagram model for interpreting propensity rules in collision induced vibrational relaxation},
        J. Chem. Phys. \textbf{74}, 4918 (1981).
\bibitem{Sharp}
        T. E. Sharp,
        \textit{Potential-energy curves for molecular hydrogen and its ions},
        Atomic Data \textbf{2}, 119 (1971).
\bibitem{WoodwardHoffmann}
        R. B. Woodward and R. Hoffmann,
        \textit{Stereochemistry of Electrocyclic Reactions},
        J. Am. Chem. Soc. 1965 \textbf{87}, 395 (1965).
\bibitem{Arranz.LiCN-HCN-HO2.correlation.diagram}
        F.~J.~Arranz, L.~Seidel, C.~G.~Giralda, R.~M.~Benito, and F.~Borondo,
        \textit{Scars at the edge of the transition from order to chaos in the isomerizing molecular systems LiNC-LiCN and HCN-HNC, and HO$_2$},
        Phys.\ Rev.\ E \textbf{82}, 026201 (2010).
\bibitem{Haake}
        F. Haake,
        \textit{Quantum Signatures of Chaos}
        (Springer Verlag, Berlin, 2010).
\bibitem{Weyl}
        H. Weyl,
        \textit{\"Uber die asymptotische verteilung der eigenwerte},
        Nachr. Konigl. Ges. Wiss. G\"ottingen, \textbf{110}, 117 (1911).
\bibitem{LL}
        A. J. Lichtenberg and M. A. Lieberman,
        \textit{Regular and Chaotic Dynamics}
        (Springer-Verlag, New York, 1992).
\bibitem{Berry}
        M. V. Berry,
        \textit{Regular and irregular motion},
        AIP Conf. Proc. \textbf{46}, 16 (1978).
\bibitem{Sibert}
        E. L. Sibert, W. P. Reinhardt, and J. T. Hynes,
        \textit{Classical dynamics of energy transfer between bonds in ABA triatomics},
        J. Chem. Phys. \textbf{77}, 3583 (1982).
\bibitem{Smale}
        S. Smale,
        \textit{Differentiable dynamical systems},
        Bull. Amer. Math. Soc. \textbf{73}, 747 (1967).
\bibitem{Chirikov}
        B. V. Chirikov,
        \textit{A universal instability of many-dimensional oscillator systems},
        Phys. Rep. \textbf{5}, 263 (1979).
\bibitem{StrattHandyMiller}
        R. M. Stratt, N. C. Handy, and W. H. Miller,
        \textit{On the quantum mechanical implications of classical ergodicity},
        J. Chem. Phys. \textbf{71}, 3311 (1979).
\bibitem{Jung}
        M. P. Jacobson, C. Jung, H. S. Taylor, and R. W. Field,
        \textit{State-by-state assignment of the bending spectrum of acetylene at 15000 cm$^{-1}$: A case study of quantum-classical correspondence},
        J. Chem. Phys. \textbf{111}, 600 (1999).
\bibitem{Shnirelman}
        A. I. Shnirelman,
        \textit{Ergodic properties of eigenfunctions},
        Ups. Mat. Nauk. \textbf{29}, 181 (1974).
\bibitem{Heller}
        E. J. Heller,
        \textit{Bound-state eigenfunctions of classically chaotic Hamiltonian systems: scars of periodic orbits},
        Phys. Rev. Lett. \textbf{53}, 1515 (1984).
\bibitem{Fabio2}
        F. Revuelta, E. G. Vergini, R. M. Benito, and F. Borondo,
        \textit{Semiclassical basis sets for the computation of molecular vibrational states},
        J. Chem. Phys. \textbf{146}, 014107 (2017).
\bibitem{Fabio3}
        F. Revuelta, E. G. Vergini, R. M. Benito, and F. Borondo,
        \textit{Scar functions, barriers for chemical reactivity, and vibrational basis sets},
        J. Phys. Chem. A \textbf{120}, 4928 (2016).
\bibitem{Polavieja}
        G. G. de Polavieja, F. Borondo, and R. M.Benito,
        \textit{Scars in groups of eigenstates in a classically chaotic system},
        Phys. Rev. Lett. \textbf{73}, 1613 (1994).
\bibitem{Sibert2}
        E. L. Sibert III, E. Vergini, R. M. Benito, and F Borondo,
        \textit{Quantum localization through interference on homoclinic and heteroclinic circuits},
        New J. Phys. \textbf{10}, 053016 (2008).
\bibitem{Fabio1}
        F. Revuelta, E. G. Vergini, R. M. Benito, and F. Borondo,
        \textit{Computationally efficient method to construct scar functions},
        Phys. Rev. E \textbf{85}, 026214 (2012).
\bibitem{Wigner}
        E. Wigner,
        \textit{On the quantum corrections for thermodynamic equilibrium},
        Phys. Rev. \textbf{40}, 749 (1932).
\bibitem{Husimi}
        K. Husimi,
        \textit{Some formal properties of the density matrix},
        Proc. Phys. Math. Soc. Jpn. \textbf{22}, 264 (1940).
\bibitem{Wisniacki}
        D. A. Wisniacki, M. Saraceno, F. J. Arranz, R. M. Benito, and F. Borondo,
        \textit{Poincar\'e-Birkhoff theorem in quantum mechanics},
        Phys. Rev. E \textbf{84}, 026206 (2011).
\bibitem{Farantos1}
        J. Tennyson and S. C. Farantos,
        \textit{Vibrational chaos in KCN - A comparison of quantum and classical calculations},
        Chem. Phys. Lett. \textbf{109}, 160 (1984).
\bibitem{Tennyson1}
        J. Tennyson and A. van der Avoird,
        \textit{Ab initio vibrational-rotational spectrum of potassium cyanide: KCN. II. Large amplitude motions and rovibrational coupling},
        J. Chem. Phys. \textbf{76}, 5710 (1982).
\bibitem{Tennyson2}
        J. R. Henderson, H. A. Lam,  and  J. Tennyson,
        \textit{Highly excited vibrational states of the KCN molecule},
        J. Chem. Soc. Faraday Trans. \textbf{88}, 85 (1992).
\bibitem{Parraga.KCN.PES}
        H.~P\'arraga, F.~J.~Arranz, R.~M.~Benito, and F.~Borondo,
        \textit{Ab initio potential energy surface for the highly nonlinear dynamics of the KCN molecule},
        J. Chem.\ Phys.\ \textbf{139}, 194304 (2013).
\bibitem{LiCN.3D}
        A. Junginger, P. L. Garcia-Muller, F. Borondo, R. M. Benito, and R. Hernandez,
        \textit{Solvated molecular dynamics of LiCN isomerization: All-atom argon solvent versus a generalized Langevin bath},
        J. Chem. Phys. \textbf{144}, 024104 (2016).
\bibitem{Ezra}
        R. M. Benito, F. Borondo, J.-H. Kim, B. G. Sumpter, and G.S. Ezra,
        \textit{Comparison of classial and Qqantum phase space structure of nonrigid molecules},
        Chem. Phys. Lett. \textbf{161}, 60 (1989).
\bibitem{Arranz.LiCN.bifurcation.diagram}
        F.~J.~Arranz, R.~M.~Benito, and F.~Borondo,
        \textit{The onset of chaos in the vibrational dynamics of LiNC/LiCN},
        J. Chem.\ Phys.\ \textbf{123}, 134305 (2005).
\bibitem{JPCA}
        H. P\'arraga, F. J. Arranz, R. M. Benito, and F. Borondo,
        \textit{Above Saddle-Point Regions of Order in a Sea of Chaos in the Vibrational Dynamics of KCN},
        J. Phys. Chem. A \textbf{122}, 3433 (2018).
\bibitem{DVR-DGB}
        Z.~Ba\u ci\'c and J.~C.~Light,
        \textit{Highly excited vibrational levels of ``floppy'' triatomic molecules: A discrete variable representation--Distributed Gaussian basis approach},
        J.\ Chem.\ Phys.\ \textbf{85}, 4594 (1986).
\bibitem{Arranz.LiCN.correlation.diagram}
        F.~J.~Arranz, F.~Borondo, and R.~M.~Benito,
        \textit{Avoided crossings, scars, and transition to chaos},
        J. Chem.\ Phys.\ \textbf{107}, 2395 (1997).
\bibitem{Lichtenberg.chaos}
        A.~J.~Lichtenberg and M.~A.~Lieberman,
        \textit{Regular and Chaotic Dynamics}
        (Springer, New York, 1992).
\bibitem{Born.adiabatic.theorem}
        M.~Born and V.~Fock,
        \textit{Beweis des Adiabatensatzes},
        Z. Phys.\ \textbf{51}, 165 (1928).
\bibitem{Fernandez.hypervirial.theorems}
        F.~M.~Fern\'andez and E.~A.~Castro,
        \textit{Hypervirial Theorems}
        (Springer, Berlin, 1987).
\bibitem{Metha.random.matrix.theory}
        M.~L.~Mehta,
        \textit{Random Matrices}
        (Elsevier, Amsterdam, 2004).
\bibitem{Marcus}
        R. A. Marcus,
        \textit{Brief Comments on Perturbation Theory of a Nonsymmetric Matrix: The GF Matrix},
        J. Phys. Chem. A \textbf{105}, 2612 (2001).
\bibitem{Schmidt.diabatic}
        F. T. Schmidt,
        \textit{Diabatic and Adiabatic Representations for Atomic Collision Problems},
        Phys. Rev. \textbf{179}, 111 (1969).
\bibitem{Landau.Zener}
        L. D. Landau,
        \textit{A Theory of Energy Transfer II},
        Phys. Z. Sowjet \textbf{2}, 46 (1932).
\bibitem{Borondo1}
        F. Borondo, A. Macias, and A. Riera,
        \textit{Asymmetry effect in the neutralization reaction H$^+$+H$^-$},
        Phys. Rev. Lett. \textbf{46}, 420 (1981).
\bibitem{Borondo2}
        F. Borondo, A. Macias, and A. Riera,
        \textit{Asymmetry effect in H$^+$+H$^-$ neutralization: Application to the $n=3$ pseudocrossing},
        Chem. Phys. Lett. \textbf{100}, 63 (1982).
\bibitem{non.crossing.rule}
        J. von Neuman and E. P. Wigner,
        \textit{\"Uber das Verhalten von Eigenwerten bei adiabatischen Prozessen},
        Phys. Z. \textbf{30}, 467 (1929).
\bibitem{footnote1}
        Note that subscripts 1 and 2, referring to stretching and bending normal modes, have been exchanged, for convenience, with respect to those in Ref.~\cite{Parraga.KCN.PES}.
\bibitem{Barker.hindered.rotor}
        J.~R.~Barker and C.~N.~Shovlin,
        \textit{An approximation for hindered rotor state energies},
        Chem.\ Phys.\ Lett. \textbf{383}, 203 (2004).
\bibitem{Arranz.LiCN.scars.edge}
        F.~J.~Arranz, F.~Borondo, and R.~M.~Benito,
        \textit{Scar formation at the edge of the chaotic region},
        Phys.\ Rev.\ Lett.\ \textbf{80}, 944 (1998).
\bibitem{McHale}
        J. L. McHale,
        \textit{Molecular Spectroscopy}
        (CRC Press, Boca Raton, (2017).
\bibitem{Oregi.Henon-Heiles.Husimi.zeros}
        I.~Oregi and F.~J.~Arranz,
        \textit{Distribution of zeros of the Husimi function in systems with degeneracy},
        Phys.\ Rev.\ E \textbf{89}, 022909 (2014).
\end{thebibliography}
\end{document}